\pdfoutput=1
\documentclass[11pt,twoside,a4paper,cmspaper,final,collab]{cms-tdr}
\def\svnVersion{737f739}\def\svnDate{2025/06/10}\def\cmsCernNoTag{CERN-EP-2024-249}\def\cmsCernDate{\today}\def\cmsMessage{Published in the Journal of High Energy Physics as \href{http://dx.doi.org/10.1007/JHEP06(2025)006}{\doi{10.1007/JHEP06(2025)006}.}}
\begin{document}\cmsNoteHeader{EXO-22-007}

\newcommand{ \PGtstar}{{\HepParticle{\PGt}{}{\ast}}\xspace}
\newcommand{\Mtstar}{\ensuremath{m_{ \PGtstar}}}
\newlength\cmsTabSkip\setlength{\cmsTabSkip}{1ex}
\newcommand{\DeepTau}{DeepTau\xspace}

\cmsNoteHeader{EXO-22-007}
\title{Search for excited tau leptons in the \texorpdfstring{$\PGt\PGt\PGg$}{tau-tau-gamma} final state in proton-proton collisions at $\texorpdfstring{\sqrt{s} = 13\TeV}{sqrt(s) = 13 TeV}$}

\author*[cern]{The CMS Collaboration}

\date{\today}

\abstract{
Results are presented for a test of the compositeness of the heaviest charged lepton, \PGt, using data collected by the CMS experiment in proton-proton collisions at a center-of-mass energy of 13\TeV at the CERN LHC. The data were collected in 2016--2018 and correspond to an integrated luminosity of 138\fbinv. This analysis searches for tau lepton pair production in which one of the tau leptons is produced in an excited state and decays to a ground state tau lepton and a photon. The event selection consists of two isolated tau lepton decay candidates and a high-energy photon. The mass of the excited tau lepton is reconstructed using the missing transverse momentum in the event, assuming the momentum of the neutrinos from each tau lepton decay are aligned with the visible decay products. No excess of events above the standard model background prediction is observed. This null result is used to set lower bounds on the excited tau lepton mass. For a compositeness scale $\Lambda$ equal to the excited tau lepton mass, excited tau leptons with masses below 4700\GeV are excluded at 95\% confidence level; for $\Lambda=10\TeV$ this exclusion is set at 2800\GeV. This is the first experimental result covering this production and decay process in the excited tau mass range above 175\GeV.
}

\hypersetup{
pdfauthor={CMS Collaboration},
pdftitle={Search for excited tau leptons in the tau+tau+gamma final state in pp collisions at sqrt(s) = 13 TeV},
pdfsubject={CMS},
pdfkeywords={CMS, beyond standard model, tau physics}
}

\maketitle

\section{Introduction}
\label{sec:intro}

Ever-increasing energy frontiers have allowed probing the compositeness of various particles at smaller and smaller length scales.
The standard model~(SM) of particle physics describes the interactions between what are assumed to be fundamental particles without substructure, the leptons and quarks, via gauge bosons which transmit the electroweak and strong forces.
The SM has been extensively tested through decades of experiments.
To date all experimental tests of the compositeness of the leptons and quarks have indicated that they are pointlike fundamental particles. 

Theoretical compositeness models~\cite{Greenberg:1974qb, Pati:1975md, Greenberg:1980ri, compositeness1, Eichten1982, Eichten:1983hw, harari, Baur90, Lane:1994qn} aim to explain why quarks and leptons have a similar generational structure and why a mass hierarchy exists amongst these particles.
These models most commonly introduce a unifying building block, called a preon, as a component of all fermions.
For lepton compositeness not to have been observed, the compositeness scale (in terms of energy) must be quite large, and thus preon exchange would be seen at LHC effective energies as a contact interaction.
One of the possible results of this contact interaction is an excited state of a lepton or a quark.
This exhibits itself as a high-mass resonance that decays into a normal lepton and a gauge boson via a gauge interaction, or into three fermions via another contact interaction.

\begin{figure}[!hb]
\centering
\includegraphics[width=0.65\textwidth]{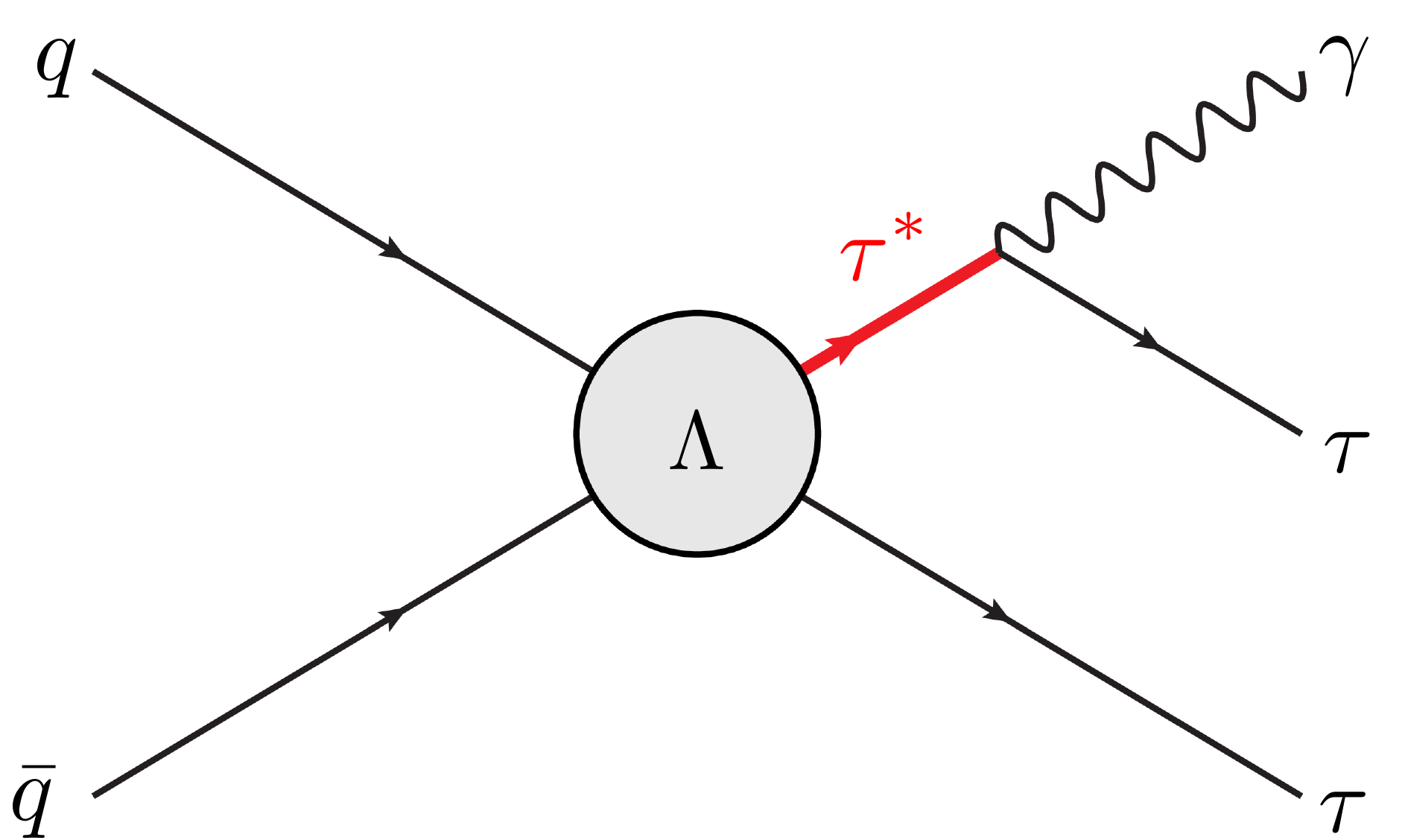}
\caption{Diagram representing the contact-interaction production of a single excited tau lepton (\PGtstar) in association with a tau lepton. The excited tau lepton decays to the ground state via the emission of a photon.}
\label{fig:vertices}
\end{figure}

This paper describes the results of an experimental test of the compositeness of the heaviest lepton, the \PGt, using data collected by the CMS experiment from proton-proton ($\Pp\Pp$) collisions at a center-of-mass energy of 13\TeV at the CERN LHC.
The data were collected in 2016--2018 and correspond to an integrated luminosity of 138\fbinv\cite{CMS:LUM-17-003, CMS:LUM-17-004, CMS:LUM-18-002}.
A search is performed seeking a tau lepton pair where one of the pair is produced in an excited state (denoted  \PGtstar), which decays to a ground state tau lepton and a photon, as depicted in Fig.~\ref{fig:vertices}.
This analysis uses the formalism presented in Ref.~\cite{Baur90}.

The Lagrangian term for  \PGtstar production via a contact interaction is given by 
\begin{equation}
    \mathcal{L}_{\mathrm{CI}} = \frac{4\pi}{2\Lambda^{2}}j^{\mu}j_{\mu}
\end{equation}
where $\Lambda$ is the compositeness scale and $j_{\mu}=\overline{q_{L}}\gamma_{\mu}q_{L}+\overline{\ell^{*}_{L}}\gamma_{\mu}\ell_{L} + \text{h.c.}$ contains left chiral quark, excited lepton, and lepton fields, respectively.
The  \PGtstar may then undergo a gauge-mediated decay governed by 
\begin{equation}
\mathcal{L}_{\mathrm{GM}} = \frac{1}{2\Lambda}\overline{\ell_{\mathrm{R}}^{*}}\sigma^{\mu\nu} \left[ gf\frac{\tau}{2}W_{\mu\nu} + g'f'\frac{Y}{2}B_{\mu\nu} \right] \ell_{\mathrm{L}} + h.c.
\label{eq:Lagrangian}
\end{equation}
where $\ell$ and $\ell^{*}$ represent SM leptons and their excited states, $W_{\PGm\PGn}$ and $B_{\PGm\PGn}$ represent the SU(2) and U(1) gauge fields, $\tau$ and $Y$ are the SU(2) and U(1) generators, and $g$ and $g'$ represent the SU(2) and U(1) gauge couplings.
The excited lepton is magnetically coupled to the lepton and the SM gauge bosons where the SU(2) and U(1) couplings are modified by the factors $f$ and $f'$, respectively, and the scale of the effective coupling is assumed to be $\Lambda$.
For simplicity and to be consistent with other searches for excited leptons, we assume $f=f'=1$.
Combining production via a contact interaction and decay via photon emission, we arrive at the focus of this analysis, a possible final state with a photon and two tau leptons, as can be seen in Fig.~\ref{fig:vertices}. 

Analogous $\Pe^{*}(\Pe\Pe\PGg)$ and $\PGm^{*}(\PGm\PGm\PGg)$ searches have been performed by CMS at $\sqrt{s} = 7,\,8$, and 13\TeV and shown no evidence of compositeness resulting in the emission of a photon \cite{CMS:2012ad, CMS:2015lgt, CMS:2018sfq}.
The most recent CMS result at 13\TeV excludes excited electrons and muons, under the assumption that the compositeness scale is equal to the excited lepton mass, below masses of 3.9 and 3.8\TeV, respectively.

A recent result from ATLAS in the $\PGt\PGt$ + 2 jet final state has set exclusions on the mass of a potential  \PGtstar~\cite{ATLAS:2023kek}; their results are complementary to ours as they assume a contact interaction decay.
The ATLAS experment has also produced results with at least three charged leptons in the final state~\cite{ATLAS:2014vih}.
The CERN LEP experiments produced reported null searches of excited leptons including \PGt leptons \cite{Buskulic:1996tw, Abreu:1998jw, Abbiendi:1999sa, Achard:2003hd}, with beam energies up to a maximum $\sqrt{s}=209\GeV$.

The search for excited taus is more challenging than that for electrons or muons as the neutrinos from the tau decays are not detectable, making it difficult to reconstruct the  \PGtstar invariant mass peak.
This analysis uses the approximation that the neutrino momentum is directed along the visible \PGt decay product direction.
This allows an accurate reconstruction of the  \PGtstar mass, as explained in Section \ref{sec:signal}.

Tabulated results are provided in the HEPData record for this analysis~\cite{hepdata}.

\section{The CMS Experiment}
\label{sec:detector}

The central feature of the CMS apparatus is a superconducting solenoid of 6 m internal diameter, providing a magnetic field of 3.8 T in the direction of the proton beam axis. Within the solenoid volume are a silicon pixel and strip tracker, a lead tungstate crystal electromagnetic calorimeter, and a brass and scintillator hadron calorimeter, each composed of a barrel and two endcap sections. Forward calorimeters extend the pseudorapidity ($\eta$) coverage provided by the barrel and endcap detectors. Muons are measured in gas-ionization detectors embedded in the steel flux-return yoke outside the solenoid. Events of interest are selected using a two-tiered trigger system. The first level, composed of custom hardware processors, uses information from the calorimeters and muon detectors to select events at a rate of around 100 kHz within a fixed latency of about 4\mus~\cite{CMS:2020cmk}. The second level, known as the high-level trigger, consists of a farm of processors running a version of the full event reconstruction software optimized for fast processing and reduces the event rate to around 1 kHz before data storage~\cite{CMS:2016ngn}. A more detailed description of the CMS detector, together with a definition of the coordinate system used and the relevant kinematic variables, can be found in Ref.~\cite{CMS:2008xjf}.

\section{Monte Carlo simulation}

Our signal model is based on a \PYTHIA~8.230 simulation \cite{Sjostrand:2014zea} using a compositeness scale of $\Lambda=10\TeV$ and  \PGtstar masses ranging from 175 to 5000\GeV.
Other small data sets were also generated for a compositenes scale $\Lambda$ equal to the mass of the  \PGtstar, and for values of $f$ and $f'$ other than 1, that showed no significant deviations in the observables described in this paper.
Therefore, all results at different compositeness scales are calculated from the limits found at $\Lambda=10\TeV$.
For \PGtstar masses close to $\Lambda$ the effective 4-fermion contact interaction may not be an adequate description of the underlying physics, depending on the details of a given model for constituent interactions. Nevertheless, excited states should not have masses much less than $\Lambda$~\cite{Baur90}.

Samples of Drell--Yan and \PW{}+jets are generated using \MGvATNLO~\cite{Alwall:2014hca} 2.6.5 generator at leading order LO using the MLM~\cite{Hoeche:2005vzu} jet matching scheme.
Samples of $\PZ\Pgg$ events are generated using \MGvATNLO at next-to-leading order and using the FxFx~\cite{Frederix:2012ps} jet matching scheme.
Samples of $\PQt\PAQt$ and single top quark events produced through $s$-channel are generated using \POWHEG 2.0~\cite{Nason:2004rx,Frixione:2007vw,Alioli:2010xd}.
Single top quark events produced through $t$-channel samples are produced using \POWHEG 2.0 in combination with \textsc{MadSpin}~\cite{Artoisenet:2012st}, a package provided as part of the \MGvATNLO framework.
The $\PW\PW$, $\PZ\PZ$, and $\PW\PZ$ samples are produced at LO using \PYTHIA 8.240.

The NNPDF3.1 next-to-next-to-leading order parton distribution function (PDF) set~\cite{Ball:2017nwa} is used for simulation of all background and signal samples.
Parton showering, fragmentation, and hadronization in all simulated samples is performed with \PYTHIA 8.240 with the CP5~\cite{CMS:2019csb} underlying event tune.
The simulation of the response of the CMS detector to incoming particles is performed using the \GEANTfour toolkit~\cite{Agostinelli:2002hh}.
Additional inelastic $\Pp\Pp$ interactions from the same or nearby bunch crossings (pileup) matching the observed distributions are simulated and incorporated in the Monte Carlo (MC) samples.

\section{Signal selection and reconstruction}
\label{sec:signal}

We seek $\Pp\Pp$ collisions that have two reconstructed \PGt decay candidates plus an isolated high-energy photon.
As a \PGt lepton may decay to an electron, muon, or hadrons, there are a total of six different final states one could potentially reconstruct.
Our signal regions (SR) comprise the three channels containing at least one hadronic \PGt decay (\tauh) and are denoted $\Pe\tauh\PGg$, $\Pgm\tauh\PGg$, and $\tauh\tauh\PGg$; final states with \Pe and \Pgm only are background dominated and have small branching fractions.
The three final states with zero \tauh are used as control regions (CR) to correct the normalization and the background simulation and associated nuisance parameters and are denoted $\Pe\Pe\PGg$, $\PGm\PGm\PGg$, and $\Pe\PGm\PGg$.
In all final states, events with additional electrons or muons are vetoed.

The reconstruction and identification of individual particles in an event is based on a particle-flow (PF) algorithm~\cite{Sirunyan:2017ulk}, which combines information from the various elements of the CMS detector.
The primary vertex (PV) is taken to be the vertex corresponding to the hardest scattering in the event, evaluated using tracking information alone, as described in Section 9.4.1 of Ref.~\cite{CMS-TDR-15-02}.
The missing transverse momentum vector \ptvecmiss is computed as the negative vector sum of the transverse momenta of all the PF candidates in an event, and its magnitude is denoted as \ptmiss~\cite{CMS:2019ctu}.
The \ptvecmiss is modified to account for corrections to the energy scale of the reconstructed jets in the event.

The reconstruction of \PGt leptons makes use of the ``hadrons-plus-strips`` algorithm, which aims to reconstruct 1 and 3-prong \tauh decays with up to two neutral pion candidates.
Identification then proceeds via a neural network named \DeepTau with classifiers to discriminate hadronically decaying \PGt leptons from jets, electrons, or muons~\cite{CMS:2022prd}.
For discrimination against hadronic jets, working points with selection efficiencies of 60, 60, and 70\% for $\Pe\tauh$, $\Pgm\tauh$, and $\tauh\tauh$ final states are used, respectively.
For discrimination against electrons, working points with selection efficiencies of 80, 99, and 99\% for $\Pe\tauh\PGg$, $\PGm\tauh\PGg$, and $\tauh\tauh\PGg$ final states are used, respectively.
For discrimination against muons, working points with selection efficiencies of 99.5\% are used for all final states.
These quoted efficiencies are derived in $\PH\to\PGt\PGt$ events for \tauh \pt in the range 30--70\GeV.
If there are more reconstructed \tauh in the event than required by the final state we select those which have the greatest \DeepTau score.

Photon identification relies on a multivariate approach that combines multiple parameters related to the photon reconstruction into a single variable to discriminate signal from background \cite{CMS:2020uim}.
Selected photons are those that pass a cut value corresponding to 90\% signal efficiency.
Additional selection criteria are applied to the photon candidate to discriminate against electrons.
In the final states that do not contain an electron, we require that there is no charged particle track that points towards the calorimeter cluster and that the tracks under consideration must not be matched to a reconstructed conversion vertex (97\% efficiency in $\PZ\PGg$ events).
In the final states containing at least one electron, we require that there is no pixel detector track seed pointing towards the calorimeter cluster and that the seed must consist of hits in at least two of the pixel detector layers (84\% efficiency in $\PZ\PGg$ events).

Candidate signal events must have been selected online by one or more triggers.
Events in the $\Pe\Pe\PGg$ and $\Pe\tauh\PGg$ final state are selected by requiring the presence of an isolated electron with $\abs{\eta}<2.5$ and $\pt > 27, 32, 32 \GeV$ in 2016, 2017, 2018, respectively.
Events in the $\PGm\PGm\PGg$ and $\PGm\tauh\PGg$ final states are selected by requiring the presence of an isolated muon with $\abs{\eta}<2.4$ and $\pt > 24, 27, 24 \GeV$ in 2016, 2017, 2018, respectively.
Events in the $\tauh\tauh\PGg$ final state are selected by requiring the presence of two \tauh with $\pt>40\GeV$ and $\abs{\eta}<2.1$.
Events in the $\Pe\PGm\PGg$ final states are selected with a logical OR of a set of triggers requiring the presence of an isolated electron, isolated muon, or both an electron and muon.
In simulation, the remaining yields are corrected using scale factors to correct for any differences in the trigger efficiency compared to that observed in data (see Section~\ref{sec:systematics}).

In the $\Pe\tauh\PGg$ and $\PGm\tauh\PGg$ final states we require at least one \tauh candidate with $\pt>20\GeV$ and $\abs{\eta}<2.3$, and photon with $\abs{\eta}<2.5$ and $\pt>100\GeV$.
In the $\tauh\tauh\PGg$ final state we require at least two \tauh candidates with $\pt>40\GeV$ and $\abs{\eta}<2.1$, and photon with $\abs{\eta}<2.5$ and $\pt>75\GeV$.
Events in the $\Pe\Pe\PGg$, $\PGm\PGm\PGg$, and $\Pe\PGm\PGg$ CRs are required to have a photon with $\abs{\eta}<2.5$ and $\pt>20\GeV$.

In the $\Pe\tauh\PGg$ and $\PGm\tauh\PGg$ final states, events having an identified \PQb quark jet with $\pt>20\GeV$ and $\abs{\eta}<2.5$ are rejected to suppress events with top quarks.
The \PQb quark jet identification uses a neural-network named ``DeepJet"~\cite{BTV-16-002, Bols:2020bkb, CMS-DP-2023-005}; we select \PQb quark jets according to a tight working point.
In the $\tauh\tauh\PGg$ final state, the expected top quark background is small and no veto on \PQb quark jets is applied.
Events in the $\Pe\tauh\PGg$, $\PGm\tauh\PGg$, and $\tauh\tauh\PGg$ final states are required to have the dilepton visible mass be larger than 100\GeV in order to reduce the Drell--Yan background.
Events in the $\Pe\Pe\PGg$ and $\PGm\PGm\PGg$ CRs are required to have the di-lepton (or $\Pell\Pell\PGg$) visible mass be within 25\GeV of the \PZ mass to select SM \PZ events, which additionally ensures neglible signal contamination.
Events in the $\Pe\PGm\PGg$ CR are required to have the dilepton visible mass be larger than 100\GeV in order to decrease the Drell--Yan contribution.
Less than 3\% of the  \PGtstar signal would appear in the $\Pe\PGm\PGg$ CR and this amount is negligible compared to the SM background.

\begin{figure}
 \centering
 \includegraphics[width=0.52\linewidth]{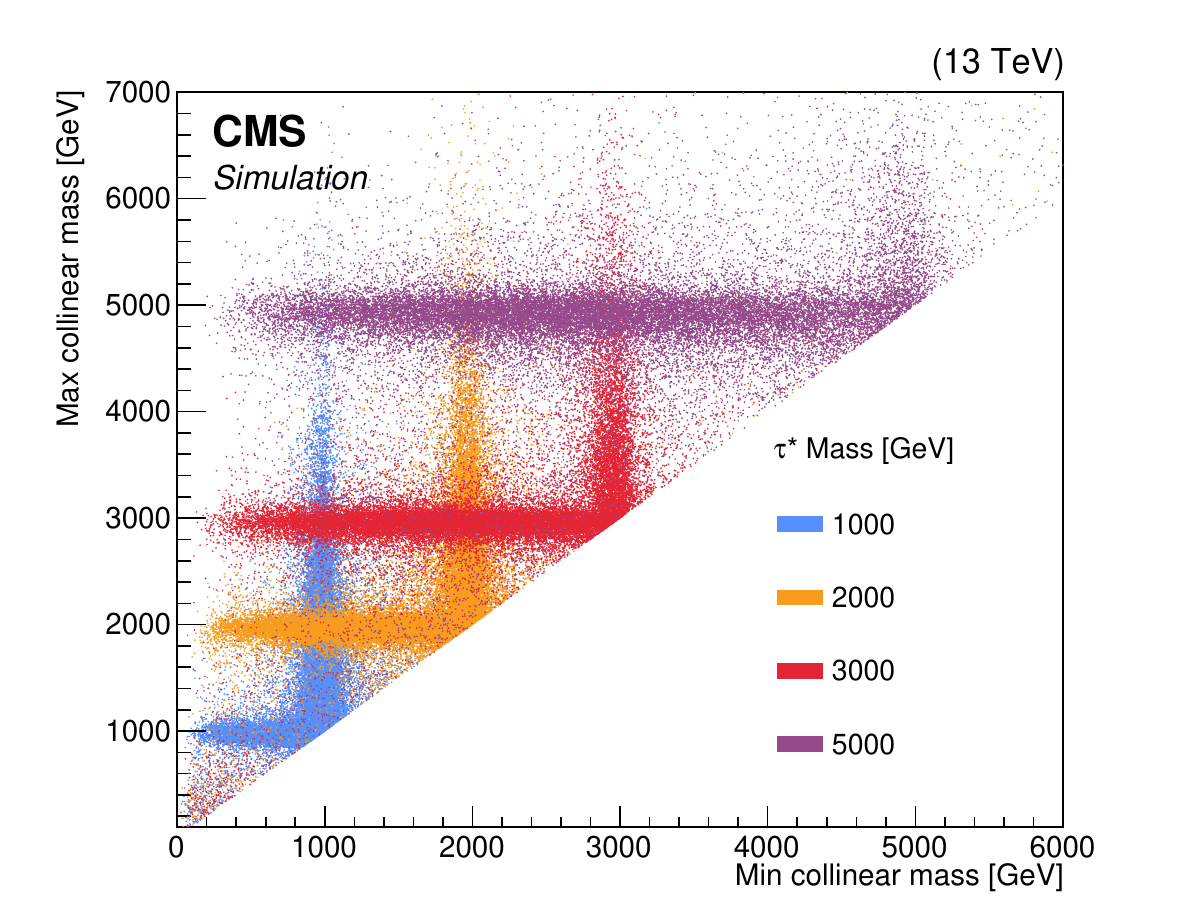}
 \includegraphics[width=0.39\linewidth]{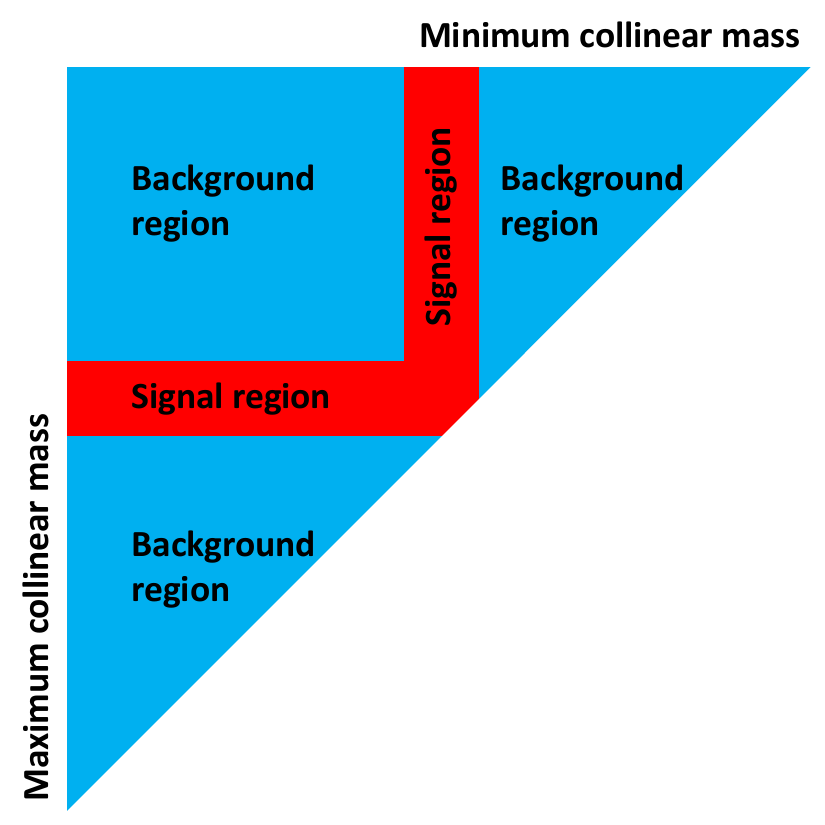}
 \caption{Left, 2-dimensional collinear mass distributions for example signal mass points. Right, a diagram showing the regions contributing to the two bins in the statistical fit.}
 \label{fig:sigmass}
\end{figure}

For a \PGt produced at high energy, the decay products tend to all lie in a narrow cone around the \PGt direction, including both the detected and undetected decay components.
If all missing transverse momentum \ptmiss is assumed to come from the tau lepton decay neutrinos, then we might approximate the momentum of the neutrinos by projecting the \ptmiss vector onto the directions of each visible \Pe, \PGm, or \tauh in the event.
This ``collinear approximation'' gives us an estimate of the initial tau lepton momentum, which we then use to reconstruct the  \PGtstar mass, in combination with the selected high-energy photon.
This approximation is highly accurate given the large boost factor of the taus considered.
As there are two tau leptons in the event, there is a twofold ambiguity as to which makes the correct pairing with the photon.
We calculate the mass of each pairing denoting them as the minimum and maximum values within an event.
The two-dimensional distribution of these masses is then used to discriminate signal from background events.
Example signal distributions are shown in Fig.~\ref{fig:sigmass}.
The correct combinations lie in an ``L''-shaped region of this distribution.
For each  \PGtstar mass hypothesis, we create a two-bin scheme in this plane corresponding to background and signal dominated regions.
The overall sensitivity of the analysis is optimized by choosing boundaries that depend on the mass of the  \PGtstar being sought and are defined by a strip centered at the hypothesis mass with a width such that the signal region contains 90\% of the given signal mass reconstructed.
Signal region boundaries are listed in Table~\ref{tab:sigBinWidths}.
An independent fit is performed for each  \PGtstar mass hypothesis.

\begin{table}[h]
\centering
\topcaption{Boundaries of the signal region for each  \PGtstar mass considered. All units are in \GeV.}
\begin{tabular}{cccc}
\hline
 \PGtstar Mass & Boundaries &  \PGtstar Mass & Boundaries\\
\hline
175  & [91.0, 259.0]     & 250  & [162.5, 337.5]\\
375  & [277.5, 472.5]    & 500  & [395.0, 605.0]\\
625  & [506.25, 743.75] & 750  & [622.5, 877.5]\\
1000 & [850.0, 1150.0]  & 1250 & [1087.5, 1412.5]\\
1500 & [1305.0, 1695.0] & 1750 & [1540.0, 1960.0]\\
2000 & [1780.0, 2220.0] & 2500 & [2250.0, 2750.0]\\
3000 & [2670.0, 3330.0] & 3500 & [3150.0, 3850.0]\\
4000 & [3560.0, 4440.0] & 4500 & [3915.0, 5085.0]\\
5000 & [4300.0, 5700.0] &      & \\
\hline
\end{tabular}
\label{tab:sigBinWidths}
\end{table}

\section{Backgrounds}
\label{sec:background}

The SM background events that are selected can be classified according to whether the \tauh candidates come from hadronic jets, genuine \PGt leptons, or (very rarely) electrons or muons.
We use MC simulation to estimate the backgrounds coming from genuine tau leptons or from electrons or muons misidentified as tau leptons.
A factorization method (``ABCD'') is used to estimate backgrounds coming from jets misreconstructed as \tauh candidates.
The primary SM processes leading to backgrounds with genuine \PGt leptons include \PZ boson production, $\PQt\PAQt$ production, and diboson production ($\PW\PW$, $\PZ\PZ$, and $\PW\PZ$).
The number of events where photon, electron, or muon candidates arise from misidentification of other objects are taken from MC simulation. 

{\tolerance=1200
In order to estimate the background from jets misidentified as \tauh candidates, the ABCD method relies on two event quantities that are nominally uncorrelated: the relative charge of the two \PGt candidates and the isolation of the \tauh candidate(s).
We denote the SR (isolated \tauh, opposite-charge) in these two quantities as A.
An isolated \tauh is one that passes the chosen \DeepTau working point, which depends on the additional energy in a cone of radius $\Delta R\equiv\sqrt{\smash[b]{(\Delta\eta)^2+(\Delta\phi)^2}}=0.5$ centered on the \tauh candidate.
\par}

Region B has opposite-charge \PGt candidates but anti-isolated \tauh, and is enriched in misidentified \tauh from hadronic jets.
In regions C and D we require same-charge \PGt candidates, where the \tauh candidates are isolated in region C and anti-isolated in region D.
The expected number of events in the SR A in which a jet is misidentified as a \tauh, is estimated from the number of observed events in B, C, and D using the expression A = BC/D.
This procedure is carried out separately for both 1- and 3-prong decays of the \tauh.
In the $\tauh\tauh\PGg$ channel, at least 1 \tauh is required to be anti-isolated.
We have checked alternative definitions, requiring one and only one \tauh to be isolated or both, and the change in the background estimation is negligible.
The presence of a photon is required in all regions.

The A and B regions are binned according to the scheme discussed in Section~\ref{sec:signal}, and thus the 1 and 3-prong B regions can be thought of as templates that are scaled by a transfer factor C/D to form the background prediction.

The B, C, and D regions contain a small fraction of events that originate from genuine \tauh decays or \Pe or \PGm misidentified as \tauh (rather than misidentified hadronic jets).
These events are estimated using simulation and do not contribute to the yields used in the prediction of the jet background.

The $\Pe\Pe\PGg$, $\PGm\PGm\PGg$, and $\Pe\PGm\PGg$ CRs are fit simultaneously with all other regions and are used to correct the normalization of the MC samples and constrain the nuisance parameters not related to \tauh reconstruction.
These CRs are binned according to the number of 30\GeV jets in the event.

Events with photons of $\pt < 75, 75$,  and 60\GeV in the $\Pe\tauh\PGg$, $\PGm\tauh\PGg$, $\tauh\tauh\PGg$ channels, respectively, are used as CRs to help correct for any observed disagreements between the predicted yields in the signal-like region A calculated from BC/D and the observed yields.
In these CRs, we perform an ABCD estimation and derive a set of $\kappa$ values such that the observed jet background in the SR A is equal to $\kappa$BC/D.
Values are calculated for both 1- and 3-prong decays of the \tauh candidates, and separately for each of the three final states, resulting in a total of six numerical values.
These CRs are fit simultaneously with the SRs and scale the jet background prediction.
The collinear mass variable is not used in this CR.

\section{Systematic uncertainties}
\label{sec:systematics}

Various systematic uncertainties resulting from imperfectly measured or modeled parameters affecting the predicted signal and/or background rates and distributions can influence the final results.
Systematic uncertainties are represented by nuisance parameters in the likelihood used to extract the final results discussed below.
These include nuisance parameters governing shapes of distributions only or governing both shapes and rates.

Table \ref{tab:systematics} shows a summary of the various sources of systematic uncertainty, the magnitude of their effect on the predicted background and signal yields in the various final state channels.
Although the analysis includes many other sources, mostly correcting for differences in reconstruction or tagging efficiencies of final state particles in simulation compared to that in data, for brevity we have included only those which have an effect of at least 1\%.
These all affect both the expected signal and backgrounds taken from simulation. The relative 1-prong \DeepTau anti-jet and trigger uncertainties are larger than those for the 3-prong case because of the larger branching fraction. The PDF uncertainties at high  \PGtstar masses, which are dominant, are large due to the uncertainties in the antiquark parton densities. The relative photon identification uncertainties are larger for signal yields than for background yields because, on average, signal photons have momenta beyond the range for which the multivariate photon identification~\cite{CMS:2020uim} was optimized.

\begin{table}[!hb]
\centering
\topcaption{Summary of systematic uncertainties that affect the expected signal and background yields by at least 1\%. Values are in \% of the inclusive yield. For the columns containing no \tauh it is the size of the effect on the background. For the columns containing \tauh, the first number is the value for the background and the second number is the value for the signal with \Mtstar=1750\GeV and $\Lambda=10\TeV$.}
\begin{tabular}{lllllll}
\hline
Source & $\Pe\Pe\PGg$ & $\PGm\PGm\PGg$ & $\Pe\PGm\PGg$ & $\Pe\tauh\PGg$ & $\PGm\tauh\PGg$ & $\tauh\tauh\PGg$ \\
\hline
Integrated Luminosity & 2.0 & 2.1 & 2.1 & 2.1, 2.1 & 2.0, 2.1 & 2.1, 2.1\\
\DeepTau anti-jet 1-prong & \NA & \NA & \NA & 3.5, 9.4 & 3.7, 8.7 & 9.0, 19\\
\DeepTau anti-jet 3-prong & \NA & \NA & \NA & 1.4, 1.5 & 1.2, 1.1 & 2.8, 2.6\\
\tauh\tauh trigger 1-prong & \NA & \NA & \NA & \NA,\NA & \NA,\NA & 8.2, 11\\
\tauh\tauh trigger 3-prong & \NA & \NA & \NA & \NA,\NA & \NA,\NA & 2.2, 1.2\\
\tauh energy scale & \NA & \NA & \NA & 0.3, 0.1 & 0.5, 0.1 & 2.0, 0.0\\
\Pe identification & 0.9 & \NA & 0.6 & 0.8, 1.3 & \NA,\NA & \NA,\NA\\
\PGg identification & 5.5 & 5.7 & 4.6 & 2.1, 14 & 2.1, 13 & 2.8, 14\\
$\Pe\PGm$ trigger efficiency & \NA & \NA & 1.1 & \NA,\NA & \NA,\NA & \NA,\NA\\
\PQb-tagging & \NA & \NA & \NA & 1.1, 0.0 & 1.0, 0.0 & \NA,\NA\\
PDF sets & \NA & \NA & \NA & \NA, 23 & \NA, 22 & \NA, 24\\
\hline
\end{tabular}
\label{tab:systematics}
\end{table}

Additional uncertainty is added to the predictions of jets misidentified as \tauh to account for two potential effects:  a) bias which may be introduced by our assumption that the $\kappa$ corrections calculated in the low photon \pt region are applicable to the nominal SR, and b) the observed year-to-year variations in the $\kappa$ values.
The size of each of the additional uncertainties is set such that the post-fit (with the SRs masked) uncertainty on that particular source is just under $100\%$.

\section{Results}
\label{sec:results}

\begin{figure}
\centering
\includegraphics[width=0.5\linewidth]{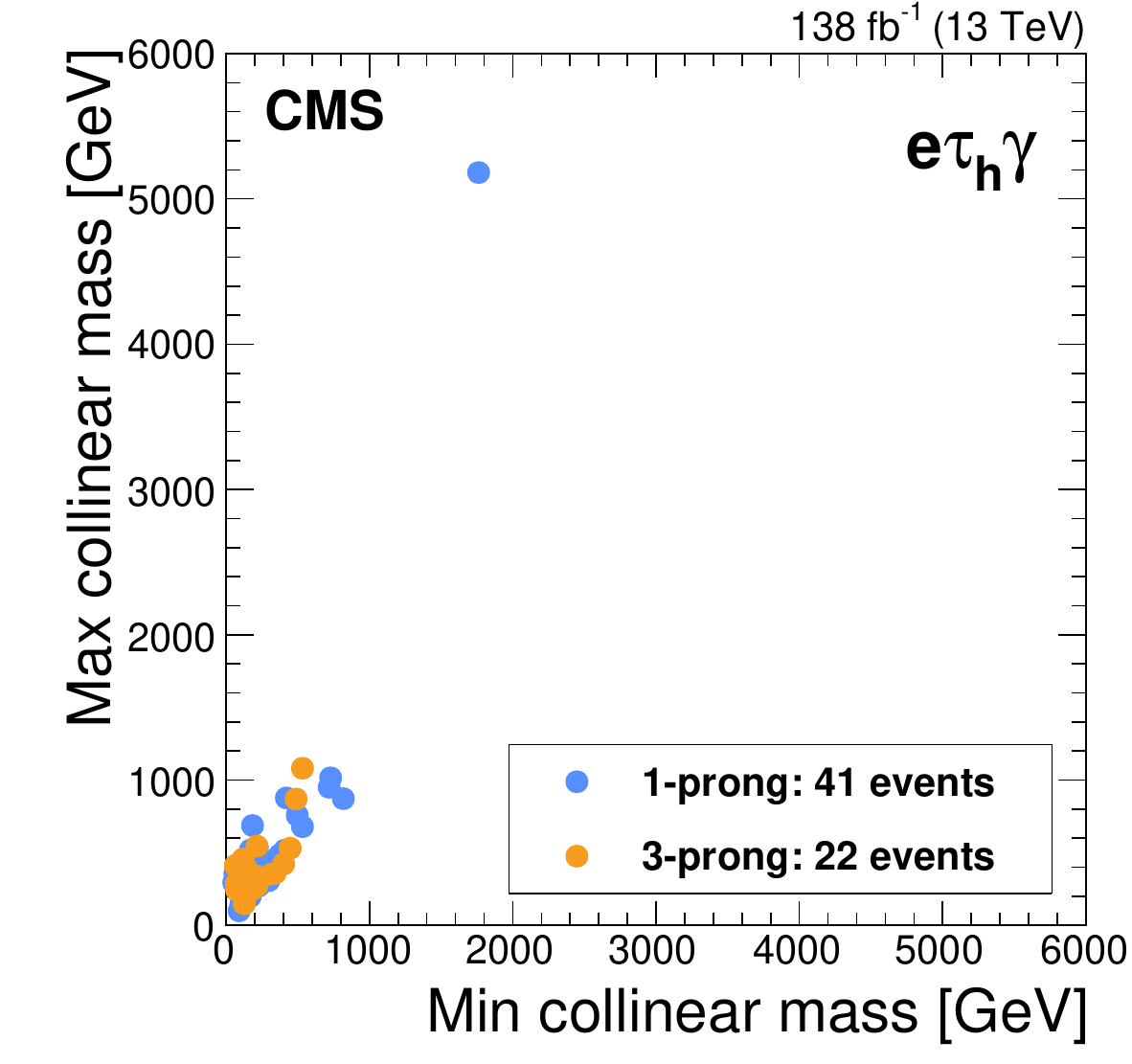}\\
\includegraphics[width=0.5\linewidth]{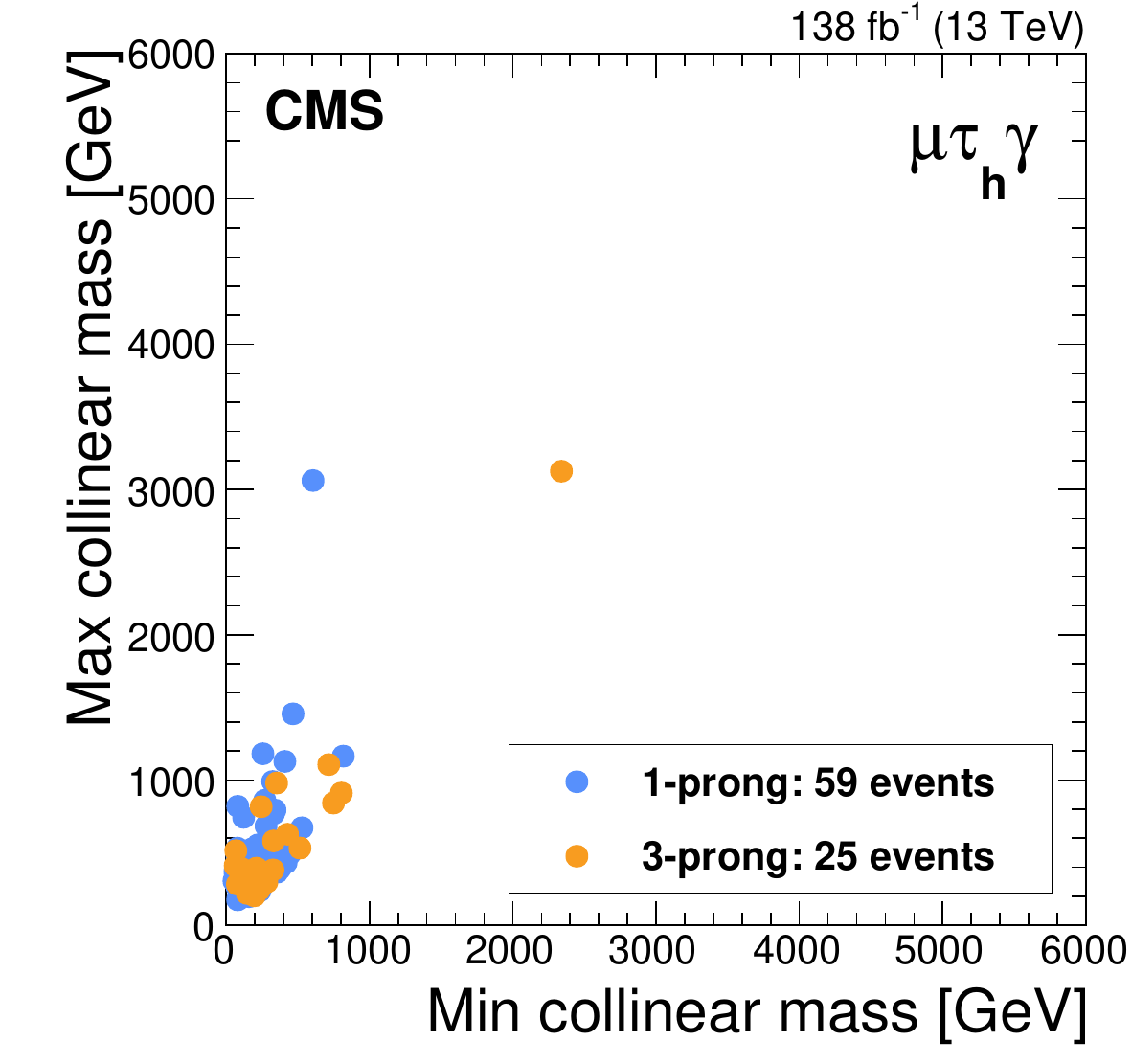}\\
\includegraphics[width=0.5\linewidth]{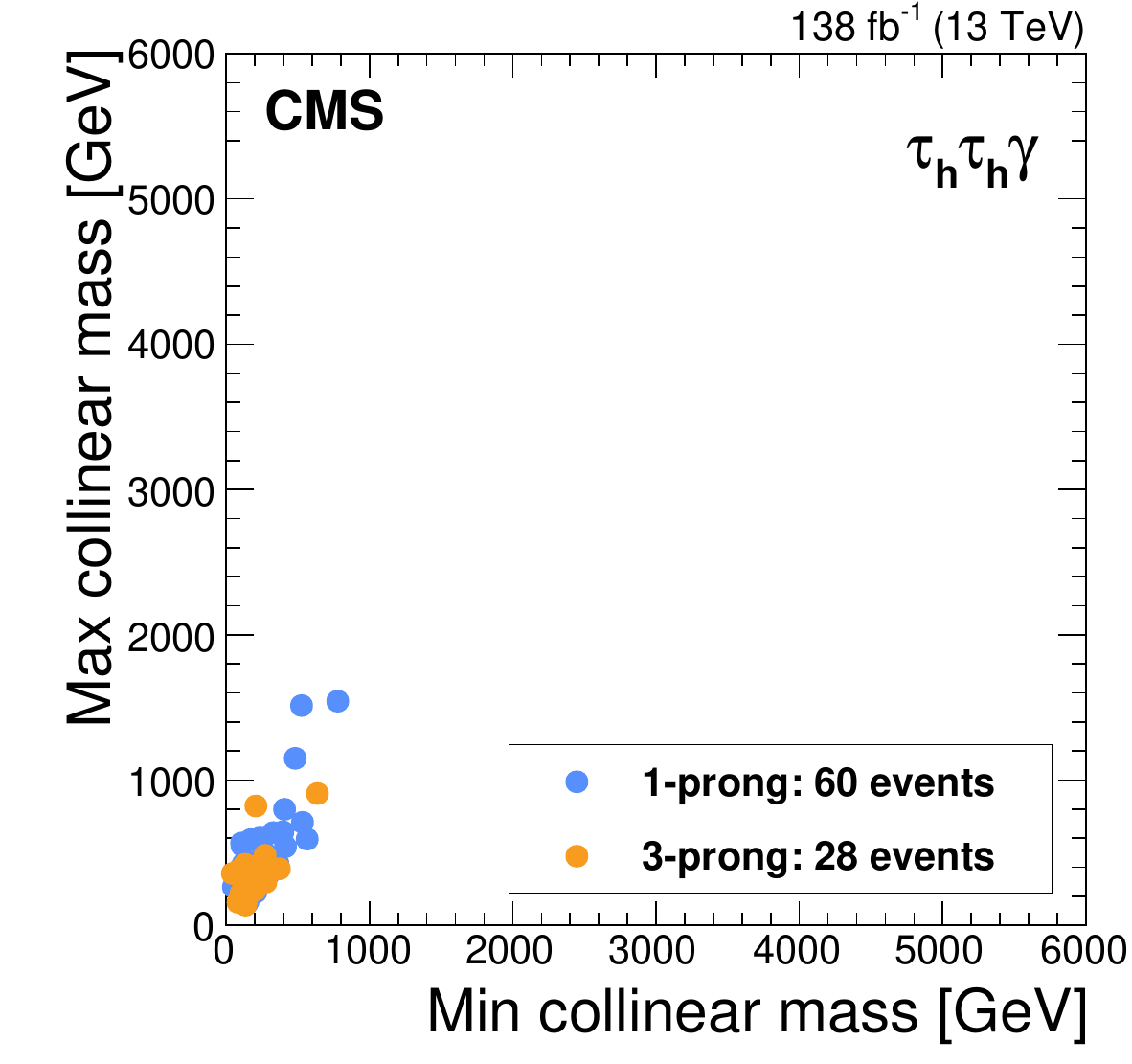}
\caption{Observed signal yields in the 2-D mass plane for the $\Pe\tauh\PGg$ (upper), $\Pgm\tauh\PGg$ (middle) and $\tauh\tauh\PGg$ (lower) final states. The three events at very high reconstructed collinear mass all have $\sin\Delta\phi$ near zero for the two \Pgt candidates, resulting in an overestimate of the mass.}
\label{fig:yields}
\end{figure}

Figure~\ref{fig:yields} shows the two-dimensional collinear mass distribution of events observed in the SR.
The statistical analysis is performed simultaneously on all signal and control regions discussed above.
Table~\ref{tab:kappa} shows the values of $\kappa$ resulting from the fit that are used to predict the jet background.
Table~\ref{tab:yields} shows the observed number of events and the expected background yields, when the analysis is not binned according to the collinear mass distribution.

\begin{table}[!h]
\centering
\topcaption{Post-fit $\kappa$ values (used in the ABCD background estimation method) calculated in the simultaneous background-only fit to all signal and control regions.}
\begin{tabular}{lccc}
\hline
         & $\Pe\tauh\PGg$ &   $\PGm\tauh\PGg$     & $\tauh\tauh\PGg$   \\
\hline
1-prong  & $1.11\pm0.13$ & $1.22\pm0.13$ & $1.09\pm0.13$ \\
3-prong  & $1.35\pm0.20$ & $1.53\pm0.25$ & $1.47\pm0.27$ \\
\hline
\end{tabular}
\label{tab:kappa}
\end{table}

\begin{table}[!h]
\centering
\topcaption{Observed events and expected background yields. The expected background yields are post-fit under the background-only hypothesis, inclusive in the collinear mass. The expected  \PGtstar yields are evaluated at $\Lambda=10\TeV$ and the uncertainties are statistical only.}
\begin{tabular}{llll}
\hline
Source & \multicolumn{1}{c}{\Pe\tauh\PGg} & \multicolumn{1}{c}{\Pgm\tauh\PGg} & \multicolumn{1}{c}{\tauh\tauh\PGg}\\
\hline
Jet (1-prong) & $25.3\pm12.8$ & $31.6\pm14.2$ & $34.4\pm16.2$ \\
Jet (3-prong) & $17.0\pm12.0$ & $14.0\pm13.1$ & $16.4\pm13.3$ \\
DY/$\PZ\PGg$ & $6.9\pm0.6$ & $16.1\pm1.1$ & $32.1\pm3.5$ \\
\ttbar & $8.5\pm0.3$ & $15.1\pm0.6$ & $4.0\pm0.3$ \\
Diboson & $3.4\pm0.5$ & $6.0\pm0.8$ & $0.2\pm0.0$ \\
Single top quark & $0.5\pm0.0$ & $0.9\pm0.1$ & $0.4\pm0.0$ \\ [\cmsTabSkip]
Total background & $61.6\pm9.7$ & $83.7\pm11.4$ & $87.5\pm11.7$ \\
Observed & \multicolumn{1}{c}{63} & \multicolumn{1}{c}{84} & \multicolumn{1}{c}{88} \\
Exp. \Mtstar=750\GeV & $35.4\pm0.5$ & $62.4\pm0.7$ & $90.5\pm0.9$\\
Exp. \Mtstar=1750\GeV & $6.1\pm0.1$ & $10.3\pm0.1$ & $15.4\pm0.2$\\
\hline
\end{tabular}
\label{tab:yields}
\end{table}

To obtain the final results in terms of the product of the cross section and branching fraction for any potential signal, we use the asymptotic approximation and the CLs criterion \cite{CLS1, CLS2, Cowan:2010js, CAT-23-001}.
This criterion is based on a binned Poisson likelihood that is a function of the signal cross section times branching fraction and predicted backgrounds.
Statistical uncertainties in the simulation are implemented with a Barlow--Beeston type method~\cite{BARLOW1993219, Conway:2011in}.
All nuisance parameters are treated as Gaussian "template morphing" parameters governing either the shape, or shape and normalization of the signal and background sources taken from simulation~\cite{CAT-23-001, Conway:2011in}.
Additional studies have been performed in which the nuisance parameters, low photon \pt CRs, and 0 \tauh CRs are removed; this has a small effect on the overall limits because the SM backgrounds are predominantly at low collinear masses, and quickly become negligible at larger  \PGtstar mass.

\begin{figure}[h]
\centering
\includegraphics[width=0.75\linewidth]{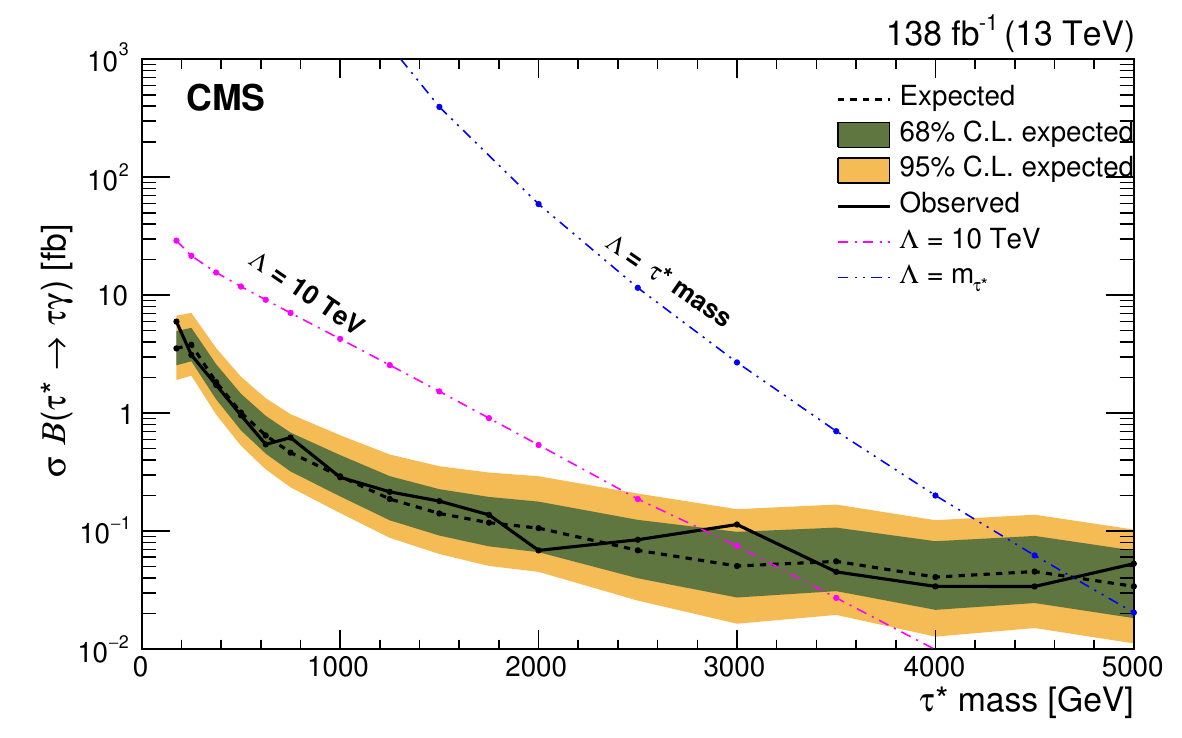}
\caption{Observed (solid-black line) and expected (dashed line) upper limits on the product of the cross section and branching fraction, as a function of the  \PGtstar mass, for single  \PGtstar production via a contact interaction in association with an SM \PGt. The inner (green) band and the outer (yellow) band indicate the regions containing 68 and 95\%, respectively, of the distribution of limits expected under the background-only hypothesis.}
\label{fig:limits}
\end{figure}

No excess above the expected background is observed.
Using these results, limits on the cross section times branching fraction have been set as a function of  \PGtstar mass and are shown in Fig.~\ref{fig:limits}.
At a compositeness scale $\Lambda$ equal to the excited  \PGtstar mass, we exclude at 95\% confidence level  \PGtstar with masses less than 4700\GeV; for $\Lambda=10\TeV$ the lower limit is set at 2800\GeV.

\section{Summary}
\label{sec:summary}
Results have been presented for a search for excited tau leptons (\PGtstar) using events with two tau leptons and a photon in the final state, produced in proton-proton collisions at a center-of-mass energy of 13\TeV.
It is assumed that the excited tau lepton is produced via a contact interaction in association with a ground state tau lepton.
The mass of the  \PGtstar is reconstructed from the missing transverse momentum, assuming the momenta of the neutrinos from the \PGt decays lie along the directions of the visible \PGt decay products.
No excess of events is observed above the expected SM backgrounds.
This null result is used to set upper limits on the cross section times branching fraction for this process.
Assuming a compositeness scale $\Lambda$ equal to the  \PGtstar mass,  \PGtstar masses less than 4700\GeV are excluded at 95\% confidence level; for $\Lambda=10\TeV$ this exclusion is set at 2800\GeV.
This is the first experimental result covering this production and decay process in the  \PGtstar mass range above 175\GeV.

\begin{acknowledgments}
We congratulate our colleagues in the CERN accelerator departments for the excellent performance of the LHC and thank the technical and administrative staffs at CERN and at other CMS institutes for their contributions to the success of the CMS effort. In addition, we gratefully acknowledge the computing centers and personnel of the Worldwide LHC Computing Grid and other centers for delivering so effectively the computing infrastructure essential to our analyses. Finally, we acknowledge the enduring support for the construction and operation of the LHC, the CMS detector, and the supporting computing infrastructure provided by the following funding agencies: SC (Armenia), BMBWF and FWF (Austria); FNRS and FWO (Belgium); CNPq, CAPES, FAPERJ, FAPERGS, and FAPESP (Brazil); MES and BNSF (Bulgaria); CERN; CAS, MoST, and NSFC (China); MINCIENCIAS (Colombia); MSES and CSF (Croatia); RIF (Cyprus); SENESCYT (Ecuador); ERC PRG, RVTT3 and MoER TK202 (Estonia); Academy of Finland, MEC, and HIP (Finland); CEA and CNRS/IN2P3 (France); SRNSF (Georgia); BMBF, DFG, and HGF (Germany); GSRI (Greece); NKFIH (Hungary); DAE and DST (India); IPM (Iran); SFI (Ireland); INFN (Italy); MSIP and NRF (Republic of Korea); MES (Latvia); LMTLT (Lithuania); MOE and UM (Malaysia); BUAP, CINVESTAV, CONACYT, LNS, SEP, and UASLP-FAI (Mexico); MOS (Montenegro); MBIE (New Zealand); PAEC (Pakistan); MES and NSC (Poland); FCT (Portugal);  MESTD (Serbia); MCIN/AEI and PCTI (Spain); MOSTR (Sri Lanka); Swiss Funding Agencies (Switzerland); MST (Taipei); MHESI and NSTDA (Thailand); TUBITAK and TENMAK (Turkey); NASU (Ukraine); STFC (United Kingdom); DOE and NSF (USA).

\hyphenation{Rachada-pisek} Individuals have received support from the Marie-Curie program and the European Research Council and Horizon 2020 Grant, contract Nos.\ 675440, 724704, 752730, 758316, 765710, 824093, 101115353, 101002207, and COST Action CA16108 (European Union); the Leventis Foundation; the Alfred P.\ Sloan Foundation; the Alexander von Humboldt Foundation; the Science Committee, project no. 22rl-037 (Armenia); the Belgian Federal Science Policy Office; the Fonds pour la Formation \`a la Recherche dans l'Industrie et dans l'Agriculture (FRIA-Belgium); the F.R.S.-FNRS and FWO (Belgium) under the ``Excellence of Science -- EOS" -- be.h project n.\ 30820817; the Beijing Municipal Science \& Technology Commission, No. Z191100007219010 and Fundamental Research Funds for the Central Universities (China); the Ministry of Education, Youth and Sports (MEYS) of the Czech Republic; the Shota Rustaveli National Science Foundation, grant FR-22-985 (Georgia); the Deutsche Forschungsgemeinschaft (DFG), among others, under Germany's Excellence Strategy -- EXC 2121 ``Quantum Universe" -- 390833306, and under project number 400140256 - GRK2497; the Hellenic Foundation for Research and Innovation (HFRI), Project Number 2288 (Greece); the Hungarian Academy of Sciences, the New National Excellence Program - \'UNKP, the NKFIH research grants K 131991, K 133046, K 138136, K 143460, K 143477, K 146913, K 146914, K 147048, 2020-2.2.1-ED-2021-00181, and TKP2021-NKTA-64 (Hungary); the Council of Science and Industrial Research, India; ICSC -- National Research Center for High Performance Computing, Big Data and Quantum Computing and FAIR -- Future Artificial Intelligence Research, funded by the NextGenerationEU program (Italy); the Latvian Council of Science; the Ministry of Education and Science, project no. 2022/WK/14, and the National Science Center, contracts Opus 2021/41/B/ST2/01369 and 2021/43/B/ST2/01552 (Poland); the Funda\c{c}\~ao para a Ci\^encia e a Tecnologia, grant CEECIND/01334/2018 (Portugal); the National Priorities Research Program by Qatar National Research Fund; MCIN/AEI/10.13039/501100011033, ERDF ``a way of making Europe", and the Programa Estatal de Fomento de la Investigaci{\'o}n Cient{\'i}fica y T{\'e}cnica de Excelencia Mar\'{\i}a de Maeztu, grant MDM-2017-0765 and Programa Severo Ochoa del Principado de Asturias (Spain); the Chulalongkorn Academic into Its 2nd Century Project Advancement Project, and the National Science, Research and Innovation Fund via the Program Management Unit for Human Resources \& Institutional Development, Research and Innovation, grant B39G670016 (Thailand); the Kavli Foundation; the Nvidia Corporation; the SuperMicro Corporation; the Welch Foundation, contract C-1845; and the Weston Havens Foundation (USA).
\end{acknowledgments}

\bibliography{auto_generated}
\cleardoublepage \appendix\section{The CMS Collaboration \label{app:collab}}\begin{sloppypar}\hyphenpenalty=5000\widowpenalty=500\clubpenalty=5000
\cmsinstitute{Yerevan Physics Institute, Yerevan, Armenia}
{\tolerance=6000
A.~Hayrapetyan, A.~Tumasyan\cmsAuthorMark{1}\cmsorcid{0009-0000-0684-6742}
\par}
\cmsinstitute{Institut f\"{u}r Hochenergiephysik, Vienna, Austria}
{\tolerance=6000
W.~Adam\cmsorcid{0000-0001-9099-4341}, J.W.~Andrejkovic, L.~Benato\cmsorcid{0000-0001-5135-7489}, T.~Bergauer\cmsorcid{0000-0002-5786-0293}, S.~Chatterjee\cmsorcid{0000-0003-2660-0349}, K.~Damanakis\cmsorcid{0000-0001-5389-2872}, M.~Dragicevic\cmsorcid{0000-0003-1967-6783}, P.S.~Hussain\cmsorcid{0000-0002-4825-5278}, M.~Jeitler\cmsAuthorMark{2}\cmsorcid{0000-0002-5141-9560}, N.~Krammer\cmsorcid{0000-0002-0548-0985}, A.~Li\cmsorcid{0000-0002-4547-116X}, D.~Liko\cmsorcid{0000-0002-3380-473X}, I.~Mikulec\cmsorcid{0000-0003-0385-2746}, J.~Schieck\cmsAuthorMark{2}\cmsorcid{0000-0002-1058-8093}, R.~Sch\"{o}fbeck\cmsAuthorMark{2}\cmsorcid{0000-0002-2332-8784}, D.~Schwarz\cmsorcid{0000-0002-3821-7331}, M.~Sonawane\cmsorcid{0000-0003-0510-7010}, W.~Waltenberger\cmsorcid{0000-0002-6215-7228}, C.-E.~Wulz\cmsAuthorMark{2}\cmsorcid{0000-0001-9226-5812}
% ...
\par}
\vskip\cmsinstskip
\dag:~Deceased\\
$^{1}$Also at Yerevan State University, Yerevan, Armenia\\
$^{2}$Also at TU Wien, Vienna, Austria\\
% ...
\end{sloppypar}
%%% END EDITABLE REGION %%%
% skeleton_end
\end{document}